\documentclass{article}

\usepackage[english]{babel}

\usepackage[letterpaper,top=2cm,bottom=2cm,left=3cm,right=3cm,marginparwidth=1.75cm]{geometry}

\usepackage{amsmath}
\usepackage{graphicx}
\usepackage[colorlinks=true, allcolors=blue]{hyperref}
\usepackage[nolist]{acronym}
\usepackage{natbib}
\usepackage{authblk}
\usepackage{float}
\usepackage{caption}
\usepackage[inline]{enumitem}
\usepackage{rotating}
\newsavebox\mysavebox
\usepackage[multiple]{footmisc}

\begin{acronym}
    \acro{EU}{European Union}
    \acro{IOM}{International Organization for Migration}
    \acro{MS}{Member State}
    \acro{OECD}{Organisation for Economic Co-operation and Development}
    \acro{SCI}{Social Connectedness Index}
    \acro{UNHCR}{United Nations High Commissioner for Refugees}
\end{acronym}

\graphicspath{ {img/} }

\title{Migration patterns, friendship networks, and the diaspora: the potential of Facebook's Social Connectedness Index to anticipate displacement patterns induced by Russia’s invasion of Ukraine in the European Union}
\author[1*]{Umberto Minora}
\author[1]{Martina Belmonte}
\author[1]{Claudio Bosco}
\author[2]{Drew Johnston}
\author[3]{Eugenia Giraudy}
\author[4]{Stefano M Iacus}
\author[1]{Francesco Sermi}
\affil[1]{Joint Research Centre, European Commission}
\affil[3]{Core Data Science, Meta}
\affil[2]{Department of Economics, Harvard University}
\affil[4]{Institute for Quantitative Social Science, Harvard University}
\affil[*]{Corresponding author: \href{mailto:umberto.minora@ec.europa.eu}{umberto.minora@ec.europa.eu}}

\date{} %% if you don't need date to appear

\begin{document}
\maketitle

\begin{abstract}
The conflict in Ukraine is causing large-scale displacement in Europe and in the World. Based on the \ac{UNHCR} estimates, more than 7 million people fled the country as of 5 September 2022. In this context, it is extremely important to anticipate where these people are moving so that national to local authorities can better manage challenges related to their reception and integration.

This work shows how innovative data from social media can provide useful insights on conflict-induced migration flows. In particular, we explore the potential of Facebook \ac{SCI} for predicting migration flows in the context of the war in Ukraine, building on previous research findings that the presence of a diaspora network is one of the major migration drivers. To do so, we first evaluate the relationship between the Ukrainian diaspora and the number of refugees from Ukraine registered for Temporary Protection or similar national schemes as a proxy of  migratory flows into the EU. We find a very strong correlation between the two (Pearson's $r=0.94$, $p<0.0001$), which indicates that the diaspora is attracting the people fleeing the war, who tend to reach their compatriots, in particular in the countries where the Ukrainian immigration was more a recent phenomenon.

Second, we compare Facebook's \acl{SCI} with available official data on diaspora at regional level in Europe. Our results suggest that the index, along with other readily available covariates, is a strong predictor of the Ukrainian diaspora at regional scale.

Finally, we discuss the potential of Facebook's \acl{SCI} to provide timely and spatially detailed information on human diaspora for those countries where this information might be missing or outdated, and to complement official statistics for fast policy response during conflicts.

\end{abstract}

\section{Introduction}
\label{sec:intro}
The war in Ukraine has caused the largest displacement of people in Europe since the World War II.\footnote{Statement by Osnat Lubrani, UN Resident \& Humanitarian Coordinator in Ukraine. \url{https://ukraine.un.org/en/175836-war-has-caused-fastest-and-largest-displacement-people-europe-world-war-ii}.} Ukraine borders four EU Member States and its citizens do not need a visa to enter the EU, being able to freely move in the Schengen area for up to 90 days \citep{2001-55-ec}.\footnote{\url{https://eu-solidarity-ukraine.ec.europa.eu/information-people-fleeing-war-ukraine_en}. Last accessed the 23 November 2022.} The response of the EU to those fleeing the war has been unprecedented  with the activation of the Temporary Protection Directive.\footnote{\url{https://home-affairs.ec.europa.eu/policies/migration-and-asylum/common-european-asylum-system/temporary-protection_en}. Last accessed the 23 November 2022.} This provides those fleeing the war with a legal status which includes an immediate access to a set of rights, including housing, education, labour. In such a context, where large movements in a free mobility area are coupled with the immediate access to services, it is important for the authorities to understand what is driving the flow from Ukraine so to anticipate the scale of people that might be arriving in a particular country. This is true for national authorities as well as for regional authorities, which have the responsibility of guaranteeing the access to the services under their competence.   

The presence of a diaspora community is one of the strongest drivers of migration. This is called ``network effect'' and is valid for labor migration as well as for movements of people seeking international protection \citep{JRC112622}. In addition, other important factors can push or pull the migratory flows. The income per capita gap between origin and destination \citep{NBERw14833}, the wage differentials between sending country and receiving country \citep{sjaastad1962costs,Harris1970}, and the share of a common language \citep{lanati2021cultural} are only some examples of such drivers. These are not included in this work since the aim is to focus solely on understanding the power of the diaspora on predicting the trajectories of the people displaced in the \ac{EU} due to the war in Ukraine.

The Ukrainian diaspora is particularly large in the \acl{EU} and even before the war Ukrainians were already among the top ten nationalities of non-EU born residents.\footnote{According to data from "Immigration by age group, sex and citizenship", accessible at \url{https://appsso.eurostat.ec.europa.eu/nui/show.do?dataset=migr_imm1ctz}.} This paper aims to understand how the Ukrainian diaspora might be driving the movement of people within the EU and what added value can  innovative data have.

The \ac{IOM} defines the diaspora as the group of 
\begin{quote}
\textit{migrants or descendants of migrants whose identity and sense of belonging, either real or symbolic, have been shaped by their migration experience and background. They maintain links with their homelands, and to each other, based on a shared sense of history, identity, or mutual experiences in the destination country.}\footnote{\ac{IOM} Glossary 2019. \url{https://publications.iom.int/system/files/pdf/iml_34_glossary.pdf}.}
\end{quote}
The diaspora of a country $X$ in a country $Y$ is often measured by the presence of people born in the country $X$ and residing in the country $Y$. However, in the case of the Ukrainian diaspora, the citizenship criterion is preferred to the criterion of the country of birth to account for the internal mobility occurred within the Soviet Union \citep{Vakhitova2020}. In line with this, we use the term ``Ukrainian diaspora'' and ``Ukrainian stocks'' to refer to Ukrainian citizens living outside Ukraine throughout the paper. This operationalisation is used in the literature \citep{BEINE201130}, however it presents some challenges. First, it does not capture short term movements of people who may not become settled migrants in the destination country but move across the origin and destination countries and have built a transitional network. This is for instance particularly relevant for Ukrainians who, thanks to their visa-free status and liberal labour migration schemes, are among the top nationalities for short-term (\textit{i.e.} less than 12 months) residence permits in the EU.\footnote{For instance, according to the migr\_resfirst series available at \url{https://ec.europa.eu/eurostat/web/migration-asylum/managed-migration/database}, they were the largest recipient of such permits in 2021.} Secondly, it does not include second generations, \textit{i.e.} people born in the country or with the citizenship of the country of residence with at least one foreign-born parent.
Thirdly, it does not capture whether the persons maintain a link with the homeland and with each other. Despite these limitations, data on the stock of foreign born or foreign citizens are to date the best available large scale comparable dataset to measure the diaspora and will be also used in this paper.

Data on the stock of migrants, however, do not always have the required spatial granularity, nor are produced with high frequency. The review of \cite{Boscoetal2022} has shown that innovative data can offer a great geographic and temporal granularity, a (near-) real time availability, and an extensive coverage suitable for more immediate international comparisons. An innovative way to measure a country's diaspora is using social media data, for instance the Facebook’s ties. In particular, Facebook's \ac{SCI} measures friendship ties of active users to assess the strength of a connection between two areas. 

In this paper, we analyse Facebook's Social Connectedness Index and traditional measures of the diaspora side-by-side. Concretely, we aim to
\begin{enumerate*}[label=(\roman*)]
    \item understand the role of the diaspora as a predictor of the displacement trajectories of those fleeing during the war in Ukraine, and
    \item test the potential of Facebook's \ac{SCI} to predict the stock of Ukrainians in the EU at detailed spatial resolution.
\end{enumerate*}

The rest of this paper is structured as follows: Section~\ref{sec:sci-diaspora-arrivals} examines the link between the actual flows of people fleeing Ukraine towards the EU and the Ukrainian diaspora in the 27 Member States; Section~\ref{sec:sci-diaspora-rel} analyses the relationship between Facebook's \ac{SCI}, and the Ukrainian stocks; finally, Section~\ref{sec:discussion_conclusion} discusses the results and concludes.

\section{Analysis of the driving power of the diaspora for the refugee flows from Ukraine}
\label{sec:sci-diaspora-arrivals}

The diaspora is one of the major drivers of migration \citep{JRC112622}. In this section, we assess the power of the diaspora in driving the movements from Ukraine to the EU. The diaspora is here measured as the stock of residents with Ukrainian citizenship.

\subsection{Description of the data}
\label{sec:data-convergence}

To quantify the movements from Ukraine we use data on the refugees from Ukraine registered for Temporary Protection or similar national protection scheme.\footnote{\url{https://data.unhcr.org/en/situations/ukraine}.} This indicator has some limitations as it does not perfectly reflect the number of Ukrainians fleeing the war who are present in the country: some people may have not registered for temporary protection (yet), hoping to return to Ukraine soon or to apply for another permit; or may have registered but moved to another country or come back to Ukraine. Yet, registrations for protection are the best proxy currently available. Data on refugees from Ukraine recorded across Europe account for people who have not registered yet; however, they present even more important limitations than the protection registrations: these data do not account for intermediate destinations, thus showing inflated counts (also known as double counts) in transit countries, especially those neighbouring Ukraine. Temporary protection data should therefore give a more likely overview of the  actual net flows, since a person who registers for temporary protection in one country shows an intention to settle there for some time, especially considering that a person cannot benefit from temporary protection in more than one country at the same time.\footnote{\url{https://eu-solidarity-ukraine.ec.europa.eu/information-people-fleeing-war-ukraine_en\#paragraph_314}.} Table~\ref{tab:tp_recorded} shows the difference between the number of temporary protections and recorded refugees.

% latex table generated in R 4.2.2 by xtable 1.8-4 package
% Fri Nov 25 13:01:17 2022
\begin{table}[H]
    \centering
    \begin{tabular}{lcc}
      \hline
    Country & Recorded refugees & Temporary Portection \\ 
      \hline
    Austria & 78158 & 78158 \\ 
      Belgium & 52245 & 52870 \\ 
      Bulgaria & 129437 & 86722 \\ 
      Croatia & 16828 & 16829 \\ 
      Cyprus & 14989 & 13113 \\ 
      Czech Republic & 412959 & 413121 \\ 
      Denmark & 31000 & 33000 \\ 
      Estonia & 32077 & 50491 \\ 
      Finland & 35240 & 35240 \\ 
      France & 96520 & 96520 \\ 
      Germany & 670000 & 971000 \\ 
      Greece & 18363 & 18363 \\ 
      Hungary & 28289 & 28289 \\ 
      Ireland & 48672 & 45074 \\ 
      Italy & 150261 & 159968 \\ 
      Latvia & 37496 & 36449 \\ 
      Lithuania & 62444 & 62444 \\ 
      Luxembourg & 6263 & 6263 \\ 
      Malta & 1284 & 1373 \\ 
      Netherlands & 68050 & 68050 \\ 
      Poland & 1274130 & 1274130 \\ 
      Portugal & 49623 & 49718 \\ 
      Romania & 52952 & 84662 \\ 
      Slovakia & 86834 & 87030 \\ 
      Slovenia & 7200 & 7200 \\ 
      Spain & 133820 & 133913 \\ 
      Sweden & 42250 & 44107 \\ 
       \hline
    \end{tabular}
    \caption{Number of recorded refugees from Ukraine across the \acl{EU} and refugees from Ukraine registered for Temporary Protection or similar national protection schemes.}
    \caption*{Source: \href{https://data.unhcr.org/en/situations/ukraine}{UNHCR}}
    \label{tab:tp_recorded}
\end{table}

To quantify the diaspora we use the number of Ukrainian citizens residing in a specific country from  Eurostat\footnote{Population on 1 January by age group, sex and citizenship, accessible at \url{https://appsso.eurostat.ec.europa.eu/nui/show.do?dataset=migr_pop1ctz&lang=en}.}, and, where not available, from UNDESA.\footnote{International Migrant Stock. \url{https://www.un.org/development/desa/pd/content/international-migrant-stock}.} These data sources provide diaspora measurements according to the country of citizenship, which is important since many Ukrainian citizens migrated to the EU from other countries that were previously part of the Soviet Union \citep{Vakhitova2020}.

Each EU country was distinguished between two classes: ``old'' diaspora countries, and more recent (``new'') diaspora countries. Old diaspora countries are defined as those where the immigration of Ukrainians used to be higher in the past, and decreased recently; new diaspora countries are those which have received a stable or increasing number of migrants from Ukraine. The reason for  distinguishing between a new and an old diaspora is that connections with the country of origin may deteriorate over time, so that people who migrated long time ago may have a smaller network in the country of origin than people who migrated only recently, and may thus be less of a driver for current movements. To distinguish between the new and the old diaspora we looked at the historical annual migration flows. We used data on long-term Ukrainian immigrants by age group, sex, and citizenship arriving into the reporting country data from Eurostat.\footnote{Immigration by age group, sex and citizenship, accessible at \url{https://appsso.eurostat.ec.europa.eu/nui/show.do?dataset=migr_imm1ctz}.} Where these data were not  available, we used the International Migration Database of the \ac{OECD}\footnote{International Migration Database, accessible at \url{https://stats.oecd.org/}.} or the residence permits issued for at least 12 months 
from Eurostat.\footnote{First permits by reason, length of validity and citizenship, accessible at \url{https://appsso.eurostat.ec.europa.eu/nui/show.do?dataset=migr_resfirst&lang=en}.} For this analysis, we selected the period 2008-2019, since 2008 is the first year for which data are complete. We excluded the year 2020 from the analysis since the COVID-19 pandemic would likely bias the final results.

\subsection{Results}

\begin{sidewaysfigure}
    \centering
    \sbox\mysavebox{\includegraphics[width=0.80\textwidth]{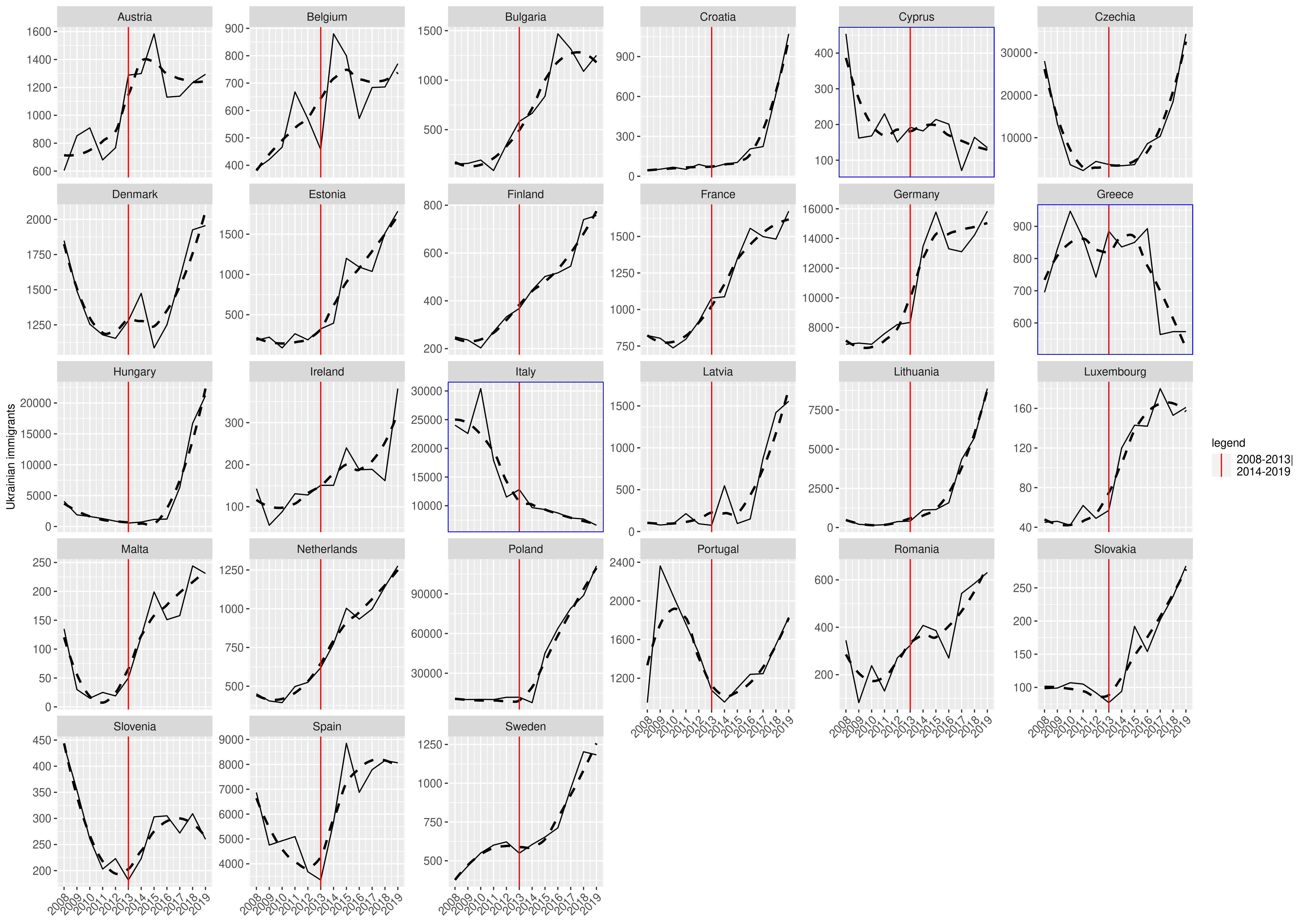}}
    \usebox\mysavebox
    \par
    \begin{minipage}{\wd\mysavebox}
        \caption{Number of Ukrainian immigrants by Member State during 2008-2019 (yearly data). The dashed line represents a smoothed version of the time series to better show the patterns. The vertical red line divides the period into two sub-periods (2008-2013, 2014-2019). Blue boxes highlight the countries classified as ``old'' diaspora.}
        \label{fig:diaspora_class}
    \end{minipage}
\end{sidewaysfigure}

We split the historical Ukrainian immigration time series on year 2013 to obtain two main periods, each six years long: 2008-2013, and 2014-2019. We then compared the number of Ukrainian immigrants in each country before and after 2013, and classified each country as having a ``new diaspora'' if \begin{enumerate*}[label=(\roman*)]
  \item the number of immigrants was greater in the 2014-2019 compared to 2008-2013, and
  \item  if the general trend of the time series was found to increase in the last period
\end{enumerate*}, otherwise they were classified as having an ``old diaspora''. Figure~\ref{fig:diaspora_class} shows the data used to classify each country.

In Figure~\ref{fig:diaspora_class} we have highlighted the countries classified as having an ``old diaspora'' with a blue box. These are Cyprus, Greece, and Italy.

To assess whether the diaspora drives the current movements from Ukraine we correlated the stock of Ukrainian citizens residing in the Member States before the war with the protection registrations from people fleeing Ukraine. The vast majority of these are Ukrainian citizens.\footnote{\url{https://appsso.eurostat.ec.europa.eu/nui/show.do?dataset=migr_asytpsm}.}
The correlation coefficient is already high considering all the countries, regardless of whether the Member State is an old or a recent destination country (Pearson's $r=0.80$, $p<0.0001$). When we focus on the countries with a new diaspora alone, the correlation coefficient increases to $0.94$.

\begin{figure}[H]
    \centering
    \includegraphics[width=0.80\textwidth]{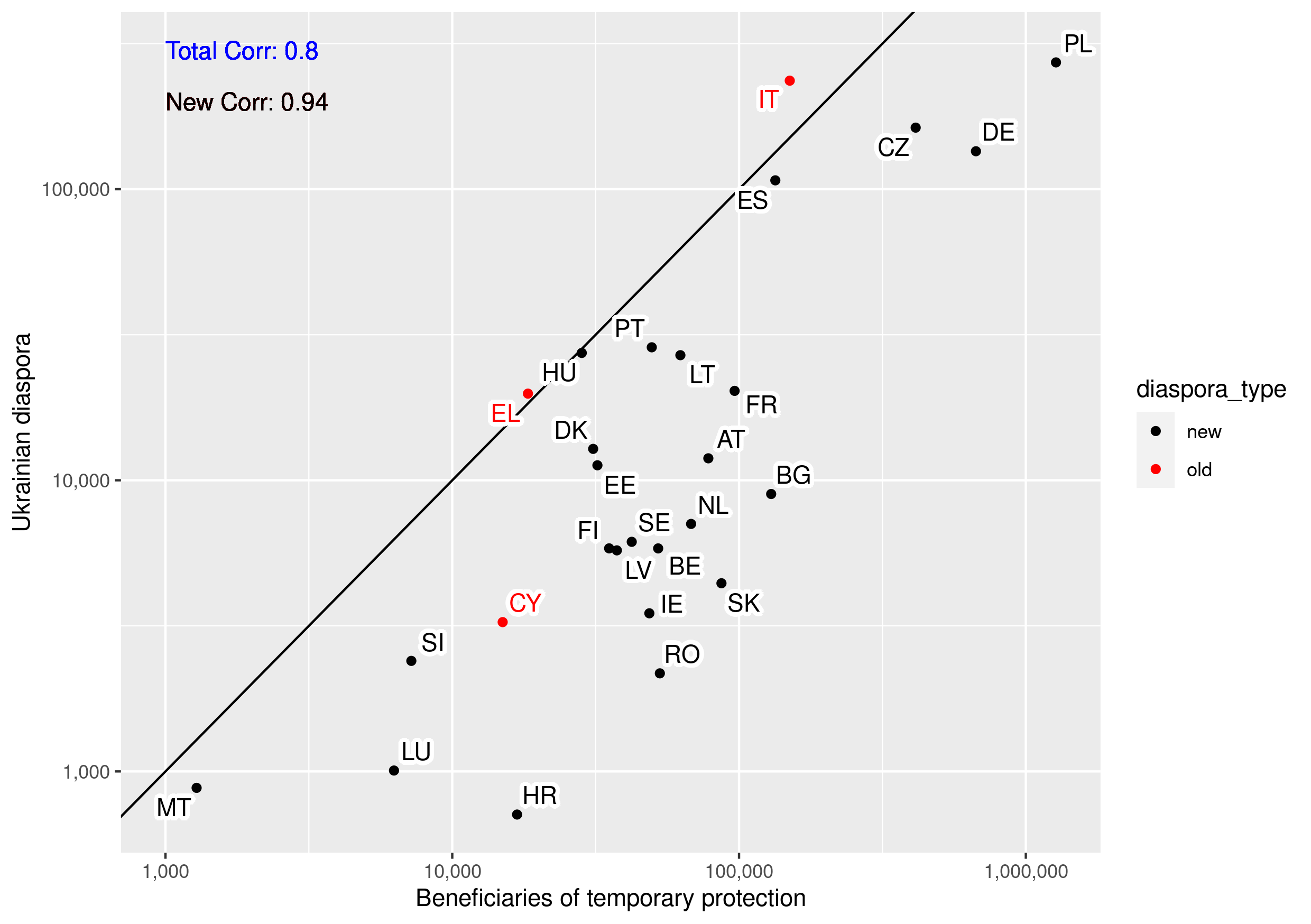}
    \caption{Number of temporary protection versus stock of Ukrainian citizens residing in the Member States before the war. The solid lines represent the bisector. Axes in logarithmic scale.}
    \label{fig:tp_diaspora}
\end{figure}

In Figure \ref{fig:tp_diaspora}, we can see that amongst the countries defined as having an ``old'' diaspora (shown in red), Italy plays an important role, since it is the country with the largest stock of Ukrainian citizens after Poland\footnote{According to data from \href{https://appsso.eurostat.ec.europa.eu/nui/show.do?dataset=migr_pop1ctz&lang=en}{EUROSTAT}, and \href{https://www.un.org/development/desa/pd/content/international-migrant-stock}{UNDESA}.}, but the ratio between its temporary protection registrations and the stock of Ukrainian citizens living there is relatively small compared to the other top five diaspora countries. This can be clearly seen in the scatter plot of Figure~\ref{fig:tp_diaspora}, where Italy lies above the bisector line whereas the rest of the countries is below. The only exception is Greece, another country classified as having an ``old diaposra'' type. This is an indication that despite having a strong historical Ukrainian diaspora, Italy and Greece seem less attractive for refugees than other countries with a more recent diaspora. This is likely related to other factors as well, such as economic, social, and political ones.

\section{Analysis of the relationship between Facebook's \acl{SCI} and the Ukrainian diaspora}
\label{sec:sci-diaspora-rel}

The findings of Section~\ref{sec:data-convergence} show that the diaspora can be used to anticipate the trajectory of displacement flows. This is true at national level and reasonably also at a more granular level. While data on residents by citizenship at regional level are provided for some Member States by the national statistical offices, we do not have a harmonised and complete dataset at this spatial granularity. In this section we therefore  assess the potential of Facebook's \ac{SCI} to fill this gap.

\subsection{Description of the data}
\label{sec:data-sci}

We collected diaspora data at regional level (\textit{i.e.} NUTS-3) from various national statistical offices, focusing on those EU countries where the Ukrainian diaspora is the largest: Italy, Spain, Germany, Czechia, and Portugal. Unfortunately, we were not able to include data for Poland, the country with the largest stock of Ukrainian citizens in the European Union, since these were not available at the same spatial resolution as the other countries and Facebook's \ac{SCI}.

We used Facebook's \ac{SCI}\footnote{\url{https://dataforgood.facebook.com/dfg/tools/social-connectedness-index}.} \citep{bailey2020determinants} before the start of the conflict in Ukraine and the total population of the hosting countries\footnote{ Population on 1 January by broad age group, sex and NUTS-3 region. \url{https://appsso.eurostat.ec.europa.eu/nui/show.do?dataset=demo_r_pjanaggr3}.} to evaluate their potential to predict human diaspora - the response variable of our analysis. Facebook's \ac{SCI} uses a snapshot of Facebook users and their friendship networks to measure the intensity of connectedness between locations in a specific time. Locations are assigned to Facebook users based on their information and activity on Facebook, including the stated city on their Facebook profile, and device and connection information. These locations do not imply a user is a citizen of that country. Facebook's \ac{SCI} between two locations $i$ and $j$ is calculated as:

\begin{equation}
\label{eq:sci}
    {Social \: Connectedness}_{ij} = \frac{{FB \: Connections}_{ij}}{{FB \: users}_i*{FB \: users}_j}
\end{equation}

${FB \: users}_i$ and ${FB \: users}_j$ are the number of Facebook users in locations $i$ and $j$, and the quantity ${FB \: Connections}_{ij}$ is the number of Facebook friendship connections between the two. The indicator  ${Social \: Connectedness}_{ij}$, therefore, is proportional to the likelihood that a given Facebook user in location $i$ is friend on Facebook with a given user in location $j$.

\begin{figure}[H]
    \centering
    \includegraphics[width=0.80\textwidth]{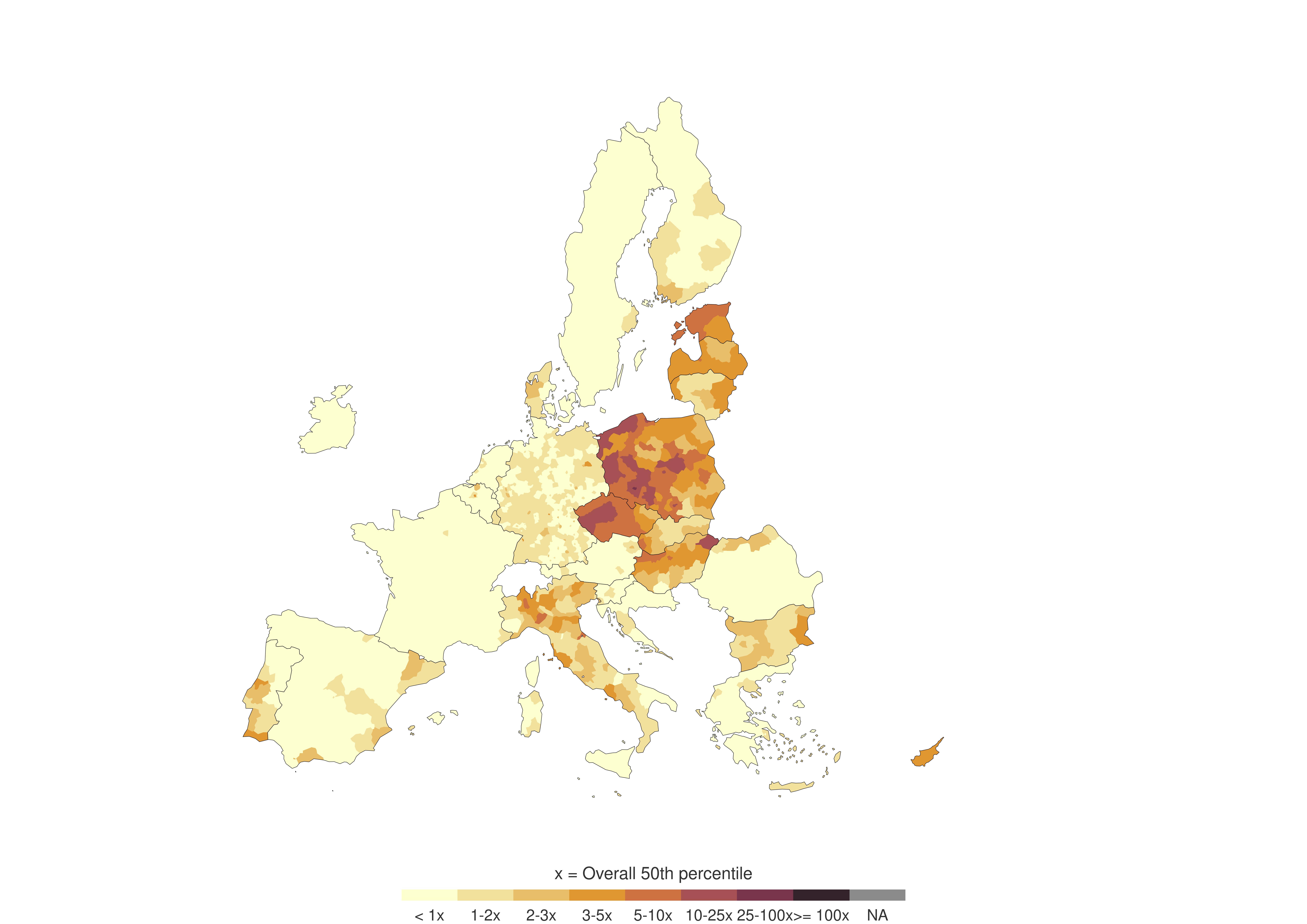}
    \caption{Facebook \acl{SCI} (Ukraine to EU27 NUTS-3). This map is for illustration purposes only. The boundaries and names shown and the designations used on this map do not imply official endorsement or acceptance by the International Organization for Migration.}
    \caption*{Source: Eurostat, \copyright EuroGeographics for the administrative boundaries}
    \label{fig:sci}
\end{figure}

Figure~\ref{fig:sci} shows the connections between Ukraine (considering the whole country), and each NUTS-3 area within the \acl{EU}.

\subsection{Results}
\label{sec:sci-analysis}

To test the hypothesis that the virtual connections (\textit{i.e.} Facebook friendship ties) between Ukraine and EU regions resemble the Ukrainian diaspora within the Union, we examined the correlation between Facebook \ac{SCI} and Ukrainian diaspora at NUTS-3 level, however the strength of this correlation varies significantly among countries (see Table~\ref{tab:mcor}). Therefore, we introduced the total population and the geographic position of each region as additional covariates in the analysis, as intuitively these are potentially linked to the diaspora. Indeed, the selected variables proved to be very good predictors for the diaspora.

\begin{table}[H] \centering 
  \caption{Correlation coefficients of Facebook \ac{SCI} and Ukrainian diaspora by country.} 
  \label{tab:mcor} 
      \begin{tabular}{@{\extracolsep{5pt}} cc} 
    \\[-1.8ex]\hline
    \hline \\[-1.8ex]
     country & corr\_coeff ($p<0.0001$) \\
    \hline \\[-1.8ex]
    CZE & $0.87$ \\
    DEU & $0.47$ \\
    ESP & $0.65$ \\
    ITA & $0.43$ \\
    PRT & $0.54$ \\
    \hline \\[-1.8ex]
    \end{tabular}
\end{table} 

\begin{figure}[H]
    \centering
    \includegraphics[width=0.80\textwidth]{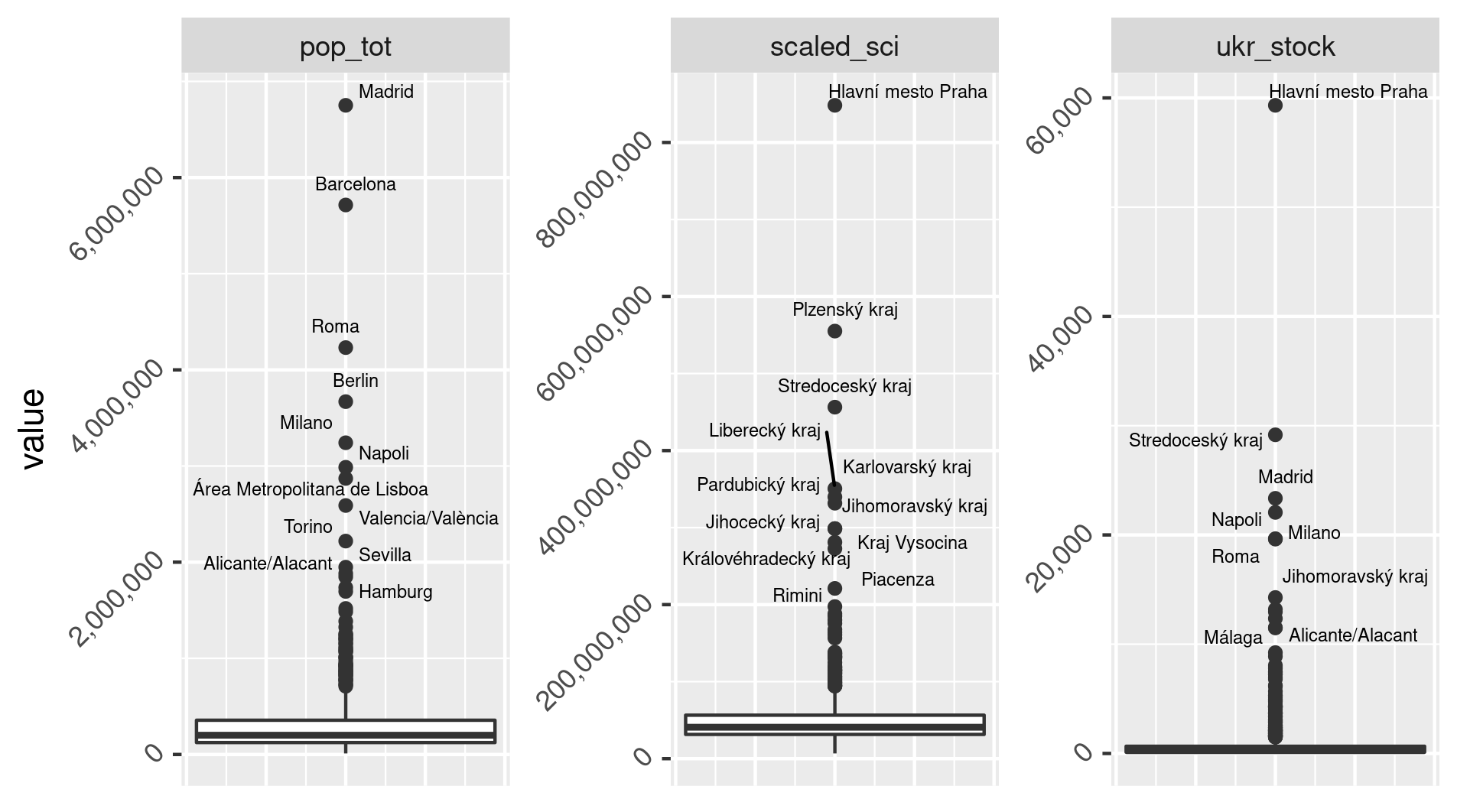}
    \caption{Distribution of total population (left), Facebook \ac{SCI} (middle), and Ukrainian diaspora (right).}
    \label{fig:distr}
\end{figure}

The Ukrainian diaspora, Facebook's \ac{SCI}, and the total population datasets all present a right-skewed distribution, meaning that each of these dataset has very few observations with very high values (compared to the mean), and these represent the capitals or other important areas (Figure~\ref{fig:distr}).

In other words, usually the most populated NUTS-3 are also those with a larger diaspora and/or a higher level of social connectedness.
Even though these regions are very few, they have a major role in driving the migratory flows. This is why we introduced a weight variable in the analysis based on the interaction between Facebook's \ac{SCI} and the total population. We derived our weights as the product between the two variables normalised between $0$ and $1$ through a min-max approach.
%Even though these regions are very few, they account for the majority of the total diaspora, and thus they are potentially more important in driving the migratory flows. It is therefore essential that our model accurately estimates them. Since in most cases a large value of diaspora corresponds to a large value of total population or Facebook \ac{SCI}, we created a weight variable based on the interaction between Facebook's \ac{SCI} and the total population. After rescaling these two variables to change their values to have a common scale ($[0,1]$) using a min-max normalization, we derived our weights as the product between the two normalized terms.

\begin{equation}
\label{eq:weight}
    W = \tilde{P}\cdot \tilde{S}
\end{equation}
In Equation~\ref{eq:weight}, $\tilde{P}$ and $\tilde{S}$ represent the normalized vectors of total population and Facebook \ac{SCI} respectively, and $W$ is the vector of the resulting weights.

Finally, we used the following model to evaluate the predictive power of the chosen variables:

\begin{equation}
\label{eq:model}
    \log(ukr\_stock) = \alpha + \beta_{1}\log(scaled\_sci) + \beta_{2}\log(pop\_tot) + \beta_{3}latitude + \epsilon
\end{equation}

\begin{figure}[H]
    \centering
    \includegraphics[width=0.80\textwidth]{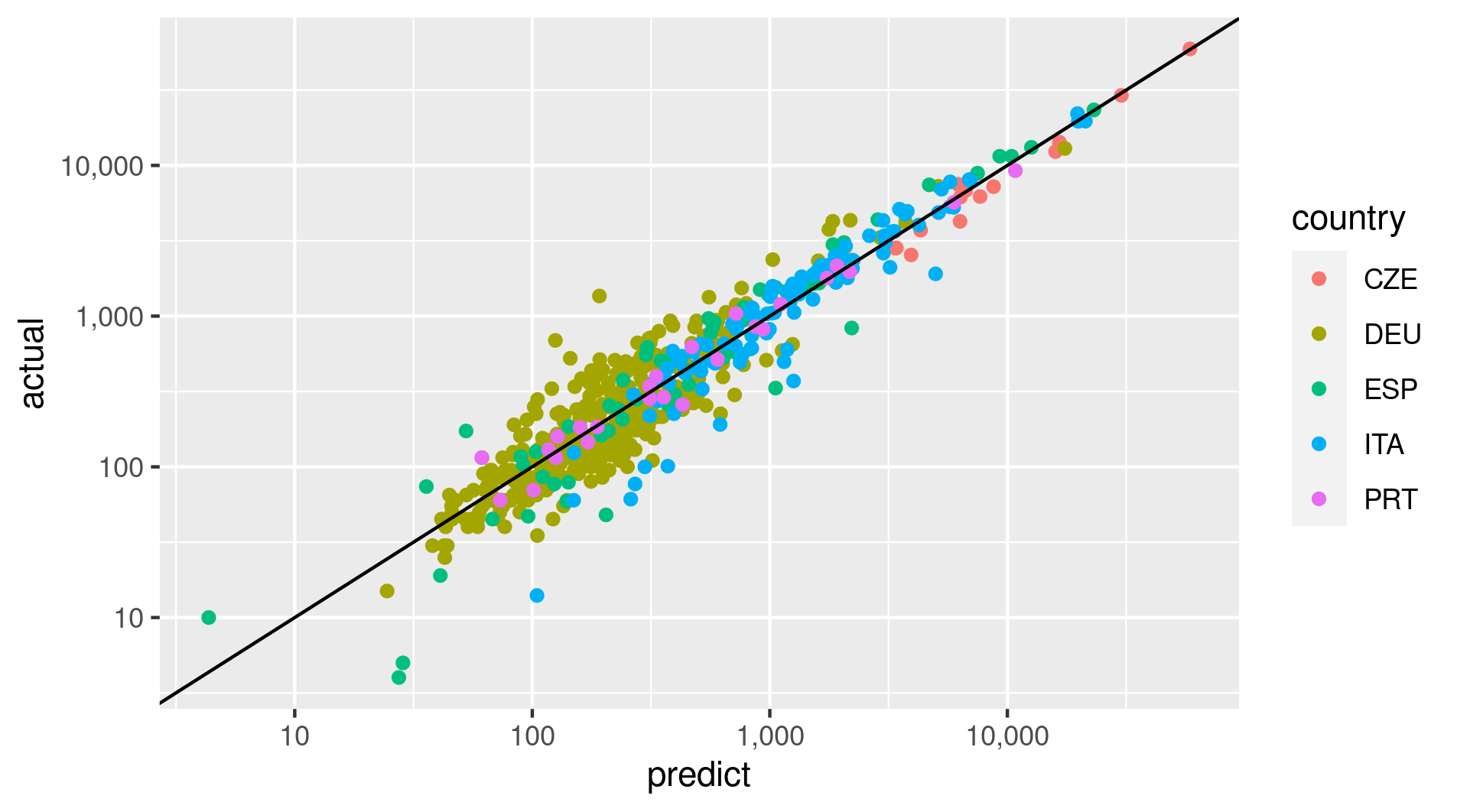}
    \caption{Scatter plot of the observed diaspora values (\textit{actual} on the $Y$ axis) versus the ones predicted by our model (\textit{pred} on the $X$ axis) coloured by country. Axes in logarithmic scale.}
    \label{fig:pred_vs_obs}
\end{figure}

We randomly selected 70\% of the total 594 observations to build a sample to train the model, and we used the remaining set to evaluate its performance. The explained variance ($R^2$, calculated on the validation set using the formula found in \cite{bosco2017exploring}) is 0.92, and all predictors have a highly significant p-value ($<0.01$). Figure~\ref{fig:pred_vs_obs} shows the scatter plot of the observed diaspora values versus the ones predicted by the model, where each color represents one of the countries available in the dataset.

\section{Discussion and conclusions}
\label{sec:discussion_conclusion}

The war in Ukraine has provoked the largest human displacement in Europe in recent years. As a response, the EU has activated the Temporary Protection Directive, which provides for immediate access to a variety of services to its beneficiaries. In this context, it becomes paramount for the authorities to be prepared and to anticipate how many people will be arriving. This is true for national as well as for local authorities, who are responsible for several of the services covered by the Directive.

The literature shows that the diaspora is a key driver of migration via a ``network effect'' \citep{JRC112622}. In this paper, we first verified the power of the diaspora to anticipate the population movements of Ukrainians following the Russian invasion. Secondly we explored the potential of innovative data, more specifically Facebook's \ac{SCI}, to measure the diaspora.

The main results of this analysis are the following.

First, the diaspora is confirmed to be an important driver of migration, especially when it is the results of migration movements that increased more recently or did not decrease over time (Section~\ref{sec:data-convergence}). The correlation between the number of Ukrainian citizens residing in EU countries (as a proxy for the diaspora) and the number of registrations for temporary protection or similar national schemes (as a proxy for migration movements) is high and significant, especially when ‘old’ migration countries are excluded. This is consistent with the hypothesis that people who migrated long time ago may have a smaller or weaker network than those who migrated more recently.

Secondly, Facebook's \ac{SCI}, along with data on total population and the geographical position of the regions, can be used to measure the diaspora at a granular spatial resolution (NUTS-3).
%In a model that includes the distance from Ukraine and the size of a country of destination, the Facebook SCI can be used to predict the distribution of the diaspora at regional level (Section~\ref{sec:sci-analysis}).
This is useful as, while the need to anticipate the arrivals is important also at regional level, spatially detalied data on the diaspora are often not available in a systematic manner. Innovative data can be used to fill this gap.

This paper paves the way for potential future research. In particular, the use of Facebook's \ac{SCI} to measure the diaspora can be further explored. Besides more granular information at spatial level, Facebook's \ac{SCI} can provide information also at high frequency in time. This can be used to more timely monitor and anticipate changes in the migratory flows occurring before the publication of official statistics.

Moreover, Facebook's \ac{SCI} can be used in complementarity with the data on migration stock to provide for a more comprehensive operationalisation of the diaspora. While widely used in the literature, the migrants stock cannot capture all the defining elements of the diaspora, in particular whether a meaningful link with the country of origin exists. Facebook's \ac{SCI} can help attest the presence of such link, although it does not tell  us whether it indicates a ``shared sense of history, identity'' and whether it has been forged by a migration experience
occurred in the current or previous generations. In addition, other important factors can push or pull the migratory flows. The income per capita gap between origin and destination, wage differentials between sending country and receiving country, and the share of a common language are only some examples of such drivers. Therefore, our future research will focus on the inclusion of the Facebook's innovative indicator in a more complex model to predict migration movements, in addition to data on the diaspora and other traditional indicators, also going beyond the Ukrainian case. 

\bibliographystyle{chicago}
\bibliography{refbib}

\end{document}